\documentclass[10pt,twocolumn,letterpaper]{article}

\usepackage{iccv}
\makeatletter
\@namedef{ver@everyshi.sty}{}
\makeatother
\usepackage{tikz}
\usepackage{times}
\usepackage{epsfig}
\usepackage{graphicx}
\usepackage{amsmath}
\usepackage{amssymb}
\usepackage{bm}
\usepackage{mathtools}
\usepackage{algorithm}
\usepackage{algpseudocode}
\usepackage{makecell}
\usepackage{multirow}
\usepackage{amsthm,thmtools}
\usepackage[accsupp]{axessibility}

\newcommand{\s}{{\boldsymbol s}}
\newcommand{\xb}{{\boldsymbol x}}
\newcommand{\yb}{{\boldsymbol y}}

\newcommand{\fb}{{\boldsymbol f}}

\definecolor{C0}{rgb}{0.121569, 0.466667, 0.705882}
\definecolor{C1}{rgb}{1.000000, 0.498039, 0.054902}
\definecolor{C2}{rgb}{0.172549, 0.627451, 0.172549}
\definecolor{C3}{rgb}{0.839216, 0.152941, 0.156863}
\definecolor{C4}{rgb}{0.580392, 0.403922, 0.741176}
\definecolor{C5}{rgb}{0.549020, 0.337255, 0.294118}
\definecolor{C6}{rgb}{0.890196, 0.466667, 0.760784}
\definecolor{C7}{rgb}{0.498039, 0.498039, 0.498039}
\definecolor{C8}{rgb}{0.737255, 0.741176, 0.133333}
\definecolor{C9}{rgb}{0.090196, 0.745098, 0.811765}
\definecolor{trolleygrey}{rgb}{0.5, 0.5, 0.5}

\usepackage[pagebackref=true,breaklinks=true,letterpaper=true,colorlinks,bookmarks=false]{hyperref}

\iccvfinalcopy 

\def\iccvPaperID{12142} 

\ificcvfinal\fi

\begin{document}

\title{Improving 3D Imaging with Pre-Trained Perpendicular 2D Diffusion Models}

\author{Suhyeon Lee\textsuperscript{\rm 1}$^{*}$, Hyungjin Chung\textsuperscript{\rm 1}$^{*}$,  Minyoung Park\textsuperscript{\rm 2}, Jonghyuk Park\textsuperscript{\rm 2}, Wi-Sun Ryu\textsuperscript{\rm 2}, Jong Chul Ye\textsuperscript{\rm 1}\\
\textsuperscript{\rm 1}Korea Advanced Institute of Science \& Technology, 
\textsuperscript{\rm 2}JLK Inc.,\\
{\tt\small \{suhyeon.lee, hj.chung, jong.ye\}@kaist.ac.kr, \{mypark, jhpark01, wsryu\}@jlkgroup.com}}


\maketitle
\def\thefootnote{*}\footnotetext{These authors contributed equally to this work}\def\thefootnote{\arabic{footnote}}
\ificcvfinal\thispagestyle{empty}\fi

\begin{figure*}[ht]
\begin{center}
\includegraphics[width=\textwidth]{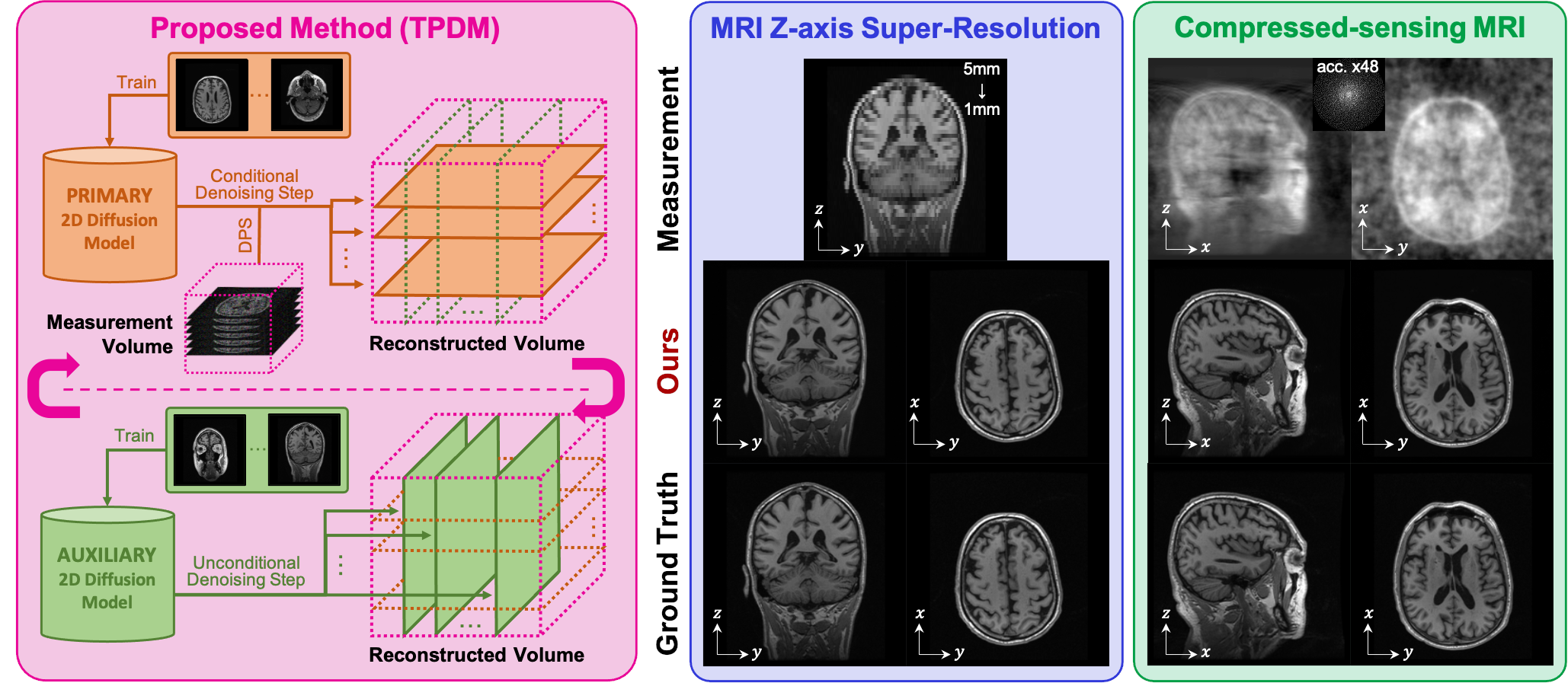}
\end{center}
   \caption{
   (Left) A visualization of our proposed method.  (Right) We display the results of solving the 3D inverse problem using the proposed method, with MR-ZSR and CS-MRI techniques shown in the center and right panels, respectively. The first row shows the measurements, the second row displays the output from our proposed method, and the third row presents the ground truth. In the MR-ZSR approach, the slice thickness was improved from 5mm to 1mm using super-resolution techniques. In the CS-MRI approach, Poisson sub-sampling was used to accelerate the process by a factor of 48.
   }
\label{fig:tpdm_title}
\end{figure*}

\begin{abstract}
Diffusion models have become a popular approach for image generation and reconstruction due to their numerous advantages. However, most diffusion-based inverse problem-solving methods only deal with 2D images, and even recently published 3D methods do not fully exploit the 3D distribution prior. To address this, we propose a novel approach using two perpendicular pre-trained 2D diffusion models to solve the 3D inverse problem. By modeling the 3D data distribution as a product of 2D distributions sliced in different directions, our method effectively addresses the curse of dimensionality. Our experimental results demonstrate that our method is highly effective for 3D medical image reconstruction tasks, including MRI Z-axis super-resolution, compressed sensing MRI, and sparse-view CT. Our method can generate high-quality voxel volumes suitable for medical applications. The code is available at \url{https://github.com/hyn2028/tpdm}
\end{abstract}

\section{Introduction}

The diffusion probabilistic model (DPM) uses neural networks to learn the gradient of the log probability distribution, $\nabla_{\bm{x}} \log p_{data}(\bm{x})$, also known as the score function. Sampling is done by either using Langevin dynamics~\cite{song2019generative} or solving the reverse stochastic differential equation (SDE) using the learned score function~\cite{song2021scorebased}.

DPM has emerged as a leading generative model in the image field since its introduction~\cite{sohl2015deep, ho2020denoising, song2021scorebased}, surpassing other models like GAN in achieving state-of-the-art performance~\cite{dhariwal2021diffusion, rombach2022high}. The confluence of the diffusion model with conditioning training is a noteworthy synergy, constituting a foundational framework within the domain of text-guided image generation~\cite{ rombach2022high, saharia2022photorealistic, ramesh2022hierarchical}. Furthermore, its versatile applicability extends to novel realms like brain vision decoding~\cite{chen2023seeing, takagi2023high}. It is also being explored as a generative model in other various areas such as audio~\cite{popov2021grad, kong2020diffwave, huang2022fastdiff}, video~\cite{blattmann2023videoldm, singer2022make, molad2023dreamix}, radiance field~\cite{muller2023diffrf, shue20233d}, and graph~\cite{vignac2022digress, huang2022graphgdp}.

Despite the slow sampling speed due to sequential sampling over multiple time steps, diffusion models offer significant advantages over other generative models, including sampling-time scalability. The pre-trained score function model can be used for conditional sampling without retraining, thanks to Bayes' theorem~\cite{dhariwal2021diffusion, ho2021classifierfree}. This conditional sampling-based inverse problem-solving method can be interpreted as posterior sampling with diffusion generative priors. Thus, it effectively avoids bias and regression to the mean phenomena from the supervised likelihood optimization methods. In due course, the paradigm of diffusion-based inverse problem-solving methodology ~\cite{song2021scorebased, kawar2022denoising, chung2022improving, chung2023diffusion, chung2022parallel, song2022solving, chung2022solving, wang2022zero} has risen to the forefront as a state-of-the-art technique within the realm of study.

Most contemporary diffusion-based inverse problem-solving methods are focused on 2D applications. However, a recent method called DiffusionMBIR \cite{chung2022solving}  has been proposed to address 3D inverse problems in medical imaging. In DiffusionMBIR, the diffusion model trained on the primary XY-plane is used as the prior, and the generative prior is augmented with a model-based prior, namely total variation (TV), to enforce smoothness to the adjacent slices (Z-axis). While this approach has been effective for various tasks, it still has limitations because it does not fully learn the 3D prior distribution of the data. More specifically, the TV prior only imposes local dependencies that are derived from finite difference operators, whereas the true 3D prior should model global dependencies.


To overcome this limitation, we propose a new method called \emph{Two Perpendicular 2D Diffusion Models (TPDM) for 3D generation}. TPDM fully leverages the 3D generative prior by modeling the 3D data distribution with a product distribution of 2D constituents, without relying on a model-based prior. This approach allows TPDM to effectively learn the 3D prior using only two 2D diffusion models: the primary model that operates on the XY-plane and an auxiliary model that learns the YZ-plane. Unlike the previous DiffusionMBIR approach, TPDM can model the global dependencies of the 3D structure, and it eliminates the need for sub-optimization schemes required to impose the TV constraint. It is worth mentioning that, unlike DiffusionMBIR which is designed specifically for inverse problem solving, TDPM is a {\em fully general 3D generative model}, which can be used both for conditional and unconditional sampling.


%

In this paper, TPDM has been tested in various 3D medical imaging reconstruction problems such as MRI Z-axis (\ie vertical axis) super-resolution (MR-ZSR), compressed sensing MRI (CS-MRI), and sparse view CT (SV-CT) and has produced the state-of-the-art results compared to existing methods.  Especially, to the best of our knowledge, we have achieved the first successful attempt at a diffusion model-based MR-ZSR 
both technically and clinically (Fig.~\ref{fig:mrzsr_result_main}). We also demonstrated that TPDM can generate a very high-quality, complete 3D voxels volume as a pure generative model (Fig.~\ref{fig:uncond_result_main}). 
Our contributions can be summarized as follows.

\vspace{-0.05cm}

\begin{enumerate} 
\item We developed a novel, simple, yet effective method to solve the 3D volume inverse problem with two perpendicular 2D diffusion models as a 3D prior, in a fully unsupervised manner, without the need for re-training.
\item We applied it to various medical imaging reconstruction problems and achieved the best-known performance. In particular, TPDM succeeded in the first attempt at a diffusion model-based MR-ZSR.  
\item Finally, we demonstrated that TPDM can also function as a 3D generative model, generating high-quality 3D voxel volumes.
\end{enumerate}

\section{Background}

\subsection{Score-based diffusion models}
The diffusion model~\cite{sohl2015deep, ho2020denoising, song2021scorebased} is a model family that defines a process that noises the original data gradually, called a forward process, and expresses the generation process by performing the learned reverse process of this noising process. Among them, the score-based diffusion model introduced by Song \etal~\cite{song2021scorebased} defines the forward process through the following Itô stochastic differential equation (SDE). Throughout the diffusion process, the data $\bm{x}$ can be represented by $\bm{x}(t)=\bm{x}_t$, with continuous time index $t\in[0, 1]$. $\bm{x}_0 \sim p_{data}$ is the raw data distribution, and $\bm{x}_1 \sim p_0$ is the predefined prior distribution.
\begin{equation} \label{forward-sde}
    d\bm{x} = \bm{f}(\bm{x},t)dt + g(t)d\bm{w},
\end{equation}
where the function $\bm{f}: \mathbb{R}^d \times \mathbb{R} \rightarrow \mathbb{R}^d$ is the drift function, and the function $g: \mathbb{R} \rightarrow \mathbb{R}$ is the diffusion coefficient. $\bm{w}$ is the standard Wiener process, also called Brownian motion.
The reverse-time SDE of Eq.~\eqref{forward-sde} can be expressed as follows~\cite{anderson1982reverse, song2021scorebased}:
\begin{equation} \label{reverse-sde}
    d\bm{x} = [\bm{f}(\bm{x}, t) - g(t)^2 \nabla_{\bm{x}_t} \log p(\bm{x}_t)] dt + g(t)d\bm{\Bar{w}}
\end{equation}
 where $\bm{\Bar{w}}$ is also the standard Wiener process.

In order to solve the reverse-time SDE for the generation process, a time-dependent score function $\nabla_{\bm{x}_t} \log p(\bm{x}_t)$ is required, which can be obtained by training the neural network-based score function estimator $\bm{s_\theta}$ through the denoising score matching (DSM) objective~\cite{vincent2011connection, song2021scorebased} 
\begin{multline} \label{denoising-score-matching}
    \min_{\bm{\theta}} \mathbb{E}_{\xb_t|\xb_0,\xb_0} [ \lVert \bm{s_\theta} (\bm{x}(t), t) - \nabla_{\bm{x}_t} \log p(\bm{x}_t | \bm{x}_0) \rVert _2^2 ]
\end{multline}
Setting $\fb(\bm{x}, t)=0$ and $g(t) = \sqrt{ \frac{d[\sigma^2(t)]} {dt} }$ with the positive time-dependent increasing noise scale function $\sigma(t)$, we achieve the so-called variance exploding SDE (VE-SDE). The sampling process of VE-SDE can be effectively solved by replacing the score function with the score network which is trained by the DSM objective. 

\subsection{Diffusion posterior sampling}
Diffusion posterior sampling (DPS) is one of the state-of-the-art methods to solve the general noisy inverse problem introduced by Chung \etal~\cite{chung2023diffusion} by using the diffusion model as a prior. 
Consider a general forward model of the inverse problem can be defined as:
\begin{equation}
    \bm{y} = \bm{A}(\bm{x}_0) + \bm{n}, \qquad \bm{y}, \bm{n} \in \mathbb{R}^n, \bm{x} \in \mathbb{R}^d
\end{equation}
where $\bm{A}$ is the forward measurement function and $\bm{n}$ is the measurement noise. To solve the inverse problem using the diffusion prior, we can use Bayes' rule to obtain
\begin{multline} \label{cond-sample-diff}
    \nabla_{\bm{x}_t} \log p(\bm{x}_t | \bm{y}) = \nabla_{\bm{x}_t} \log p(\bm{x}_t) + \nabla_{\bm{x}_t} \log p(\bm{y} | \bm{x}_t) \\
                                                \simeq \bm{s}_{\bm{\theta}^*}(\bm{x}_t, t) + \nabla_{\bm{x}_t} \log p(\bm{y} | \bm{x}_t).
\end{multline}
Nonetheless, since there is no explicit relationship between $\xb_t$ and $\yb$, we cannot use \eqref{cond-sample-diff} directly. To circumvent this problem, \cite{chung2023diffusion} proposes an approximation with a theoretically guaranteed upper bound on the approximation error
\begin{align}
    \label{dps-main}
    \nabla_{\bm{x}_t} \log p(\bm{y}|\bm{x}_t) \simeq \nabla_{\bm{x}_t} \log p(\bm{y} |\hat{\bm{x}}_0 (\bm{x}_t)),
\end{align}
where
\begin{align}
\label{dps-twee}
    \hat{\bm{x}}_0(\bm{x}_t) \coloneqq \mathbb{E}[\bm{x}_0|\bm{x}_t] = \bm{x}_t + \sigma^2(t) \nabla_{\bm{x}_t} \log p(\bm{x}_t)
\end{align}
is the Tweedie denoised estimate~\cite{efron2011tweedie, kim2021noise2score}. Accordingly, when the measurement noise is Gaussian, one can use:
\begin{equation}
        \nabla_{\bm{x}_t} \log p(\bm{x}_t | \bm{y}) \simeq  \bm{s_{\theta^*}}(\bm{x}_t, t) - \lambda \nabla_{\bm{x}_t} \lVert \bm{A}(\Hat{\bm{x}}_0(\bm{x}_t)) - \bm{y} \rVert ^2 _2.
\end{equation}

\section{Two Perpendicular 2D Diffusion Model}

\subsection{Modeling data distribution} 

To overcome the drawbacks of DiffusionMBIR, here we describe our method of applying priors that is closer to the actual 3D distribution than DiffusionMBIR. Our simple yet effective solution is, by modeling the 3D data distribution as the product distribution, to additionally use an auxiliary diffusion model trained on 2D slices in different directions of the volume, in addition to the primary 2D diffusion model to solve the inverse problem (Fig. \ref{fig:tpdm_title}). This allows us to effectively drive a diffusion model in high dimensional space, much like the utilization of factorization methods in diverse deep learning scenarios for efficiency~\cite{sainath2013low, guo2017deepfm, garbin2021fastnerf}.

Specifically, our proposal is to model the data distribution as the {\em product} distribution given by
\begin{align}
\label{eq:prod}
p_{\bm{\theta}, \bm{\phi}}(\xb) = q_{\bm{\theta}}^{(p)}(\xb)^\alpha q_{\bm{\phi}}^{(a)}(\xb)^{\beta}/Z \qquad\\
\begin{aligned}
=  [q_{\bm{\theta}}^{(p)}(\xb_{[:,:,1]}) q_{\bm{\theta}}^{(p)}(\xb_{[:,:,2]}) \cdots q_{\bm{\theta}}^{(p)}(\xb_{[:,:,d_3]})]^\alpha \\ \times[q_{\bm{\phi}}^{(a)}(\xb_{[1,:,:]}) q_{\bm{\phi}}^{(a)}(\xb_{[2,:,:]}) \cdots 
q_{\bm{\phi}}^{(a)}(\xb_{[d_1,:,:]})]^{\beta}/Z,
\end{aligned}
\end{align}
where $Z$ is an appropriate normalizing partition function,
$q_{\bm{\theta}}^{(p)}(\xb)$ is the distribution modeled by the primary model parameterized with $\bm{\theta}$, and $q_{\bm{\phi}}^{(a)}(\xb)$ is the distribution modeled by the auxiliary model parameterized with $\bm{\phi}$, for $\xb \in \mathbb{R}^{d_1 \times d_2 \times d_3}$. Moreover, $\alpha, \beta$ induces weighting between the two distributions according to the importance. 
We further assume that both $q_{\bm{\theta}}^{(p)}$ and $q_{\bm{\phi}}^{(a)}$ can be decomposed into independent 2D (slice) distributions.

Accordingly, when performing unconditional sampling from the prior distribution $p_{\bm{\theta}, \bm{\phi}}(\xb)$, we can directly use
\begin{multline}
    \nabla_{\xb_t} \log p(\xb_t) = \alpha \nabla_{\xb_t} \log q^{(p)}(\xb_t) + \beta \nabla_{\xb_t} \log q^{(a)}(\xb_t) \\
    = \alpha \Sigma_{i=1}^{d_3} \nabla_{\xb_t} \log q^{(p)}(\xb_{t, [:,:,i]}) + \beta \Sigma_{i=1}^{d_1} \nabla_{\xb_t} \log q^{(a)}(\xb_{t, [i,:,:]}) \\
    \simeq \alpha \Sigma_{i=1}^{d_3} \s^{3D}_{{\bm{\theta}}^*}(\xb_{t, [:,:,i]}) + \beta \Sigma_{i=1}^{d_1} \s^{3D}_{{\bm{\phi}}^*}(\xb_{t, [i,:,:]}),
\end{multline}
where $\xb_{t, [i,:,:]}$ and  $\xb_{t, [:,:,j]}$ denote the $i$ and $j$-th  $x$- and $z$-slice of $\xb_t$, respectively, and  
\begin{align}
\begin{cases}
    \s^{3D}(\xb_{t, [:,:,i]})_{[:,:,i]} = \s(\xb_{t, [:,:,i]}) \\
    \s^{3D}(\xb_{t, [:,:,i]})_{[\text{otherwise}]} = 0
\end{cases} \\
\begin{cases}
    \s^{3D}(\xb_{t, [i,:,:]})_{[i,:,:]} = \s(\xb_{t, [i,:,:]}) \\
    \s^{3D}(\xb_{t, [i,:,:]})_{[\text{otherwise}]} = 0
\end{cases}
\end{align}
 which used the trained 2D score estimator $\bm{s}(\cdot)$ 
 due to our 2D slice independence assumption. However, care must be taken since simply using this approximation would be compute-heavy, as one would have to evaluate two forward passes {\em per} each iteration. In this regard, we propose a simple fix to this problem by using alternating updates
\begin{equation}
\begin{aligned}
\begin{cases}
\Sigma \s^{3D}_{{\bm{\theta}}^*}(\xb_{t, [:,:,i]}), &\text{\footnotesize with $\mathbb{P} = \alpha / (\alpha + \beta)$}\\
\Sigma \s^{3D}_{{\bm{\phi}}^*}(\xb_{t, [i,:,:]}), &\text{\footnotesize with $\mathbb{P} = \beta / (\alpha + \beta)$}
\end{cases}
\end{aligned}
\label{eq:alternating_uncond}
\end{equation}
where $\mathbb{P}$ denotes the probability of each step to be performed. \eqref{eq:alternating_uncond} can be implemented in regularly structured intervals or in a stochastic fashion, which we discuss in detail in Section~\ref{solving-3d-inverse-problem-with-tpdm}.

Finally, in order to solve the inverse problem, we can leverage the following result:
\begin{equation}
\begin{aligned}
    &\nabla_{\xb_t} \log p(\xb_t|\yb) \simeq \alpha \nabla_{\xb_t} \log q^{(p)}(\xb_t)\\ &+ \beta \nabla_{\xb_t} \log q^{(a)}(\xb_t) + \nabla_{\xb_t} \log p(\yb|\hat\xb_0 (\xb_t) ),
\end{aligned}
\end{equation}
where is simplified 
 similar to unconditional sampling in \eqref{eq:alternating_uncond} as
 \begin{equation}
\begin{aligned}
\begin{cases}
\Sigma \s^{3D}_{{\bm{\theta}}^*}(\xb_{t, [:,:,i]}) + \gamma_t \nabla_{\xb_t} \log p(\yb|\hat\xb_0(\xb_t)), &\text{\footnotesize with $\mathbb{P} = \alpha / (\alpha + \beta)$}\\
\Sigma \s^{3D}_{{\bm{\phi}}^*}(\xb_{t, [i,:,:]}), &\text{\footnotesize with $\mathbb{P} = \beta / (\alpha + \beta)$}
\end{cases}
\end{aligned}
\label{eq:alternating_cond}
\end{equation}
where $\gamma_t$ is the step size that also absorbs the weighting factor induced by $\alpha$ and $\beta$.

 \subsection{Solving 3D reconstruction problem with TPDM} \label{solving-3d-inverse-problem-with-tpdm}

Training of TPDM is performed by training the primary and auxiliary 2D diffusion model (for the algorithm, see Appendix \ref{training-tpdm}). The primary 2D diffusion model $\bm{s}_{{\bm{\theta}}^*}$ selects an appropriate plane when solving the inverse problem and is trained with sliced images of 3D volumes into the corresponding plane. For example, in the case of CS-MRI and SV-CT, it is the axial plane, and in the case of MR-ZSR, it is the sagittal or coronal plane. An auxiliary 2D diffusion model $\bm{s}_{{\bm{\phi}}^*}$  is trained by selecting one of the two remaining planes of the volumes.

In order to solve the inverse problem, conditional sampling is performed alternately using the trained TPDM for each step of one time-step denoising (Algorithm~\ref{alg:inference_tpdm}). While the algorithms are presented individually for clarity, they can be batched for computational efficiency.
 In each denoising step, we use the primary diffusion model $\bm{s}_{{\bm{\theta}}^*}$ to constrain the consistency of measurements $\bm{y}$ and sample an image using the DPS~\cite{chung2023diffusion}. The hyperparameter $\lambda$ controls the strength of the measurement consistency. The auxiliary diffusion model $\bm{s}_{{\bm{\phi}}^*}$ is used to correct inconsistencies in the batch direction caused by the primary diffusion model. We adjust the contribution of the two models using the integer hyperparameter $K$ (for non-integer values of $K$, see Appendix~\ref{real-value-k}). For example, if $K$=4, the primary model and the auxiliary model contribute to image generation at a ratio of 3:1, respectively.

%
%
%

\begin{algorithm}
\caption{Solving 3D Inverse Problem with TPDM}\label{alg:inference_tpdm}
\begin{algorithmic}
\Require $\bm{Y} \in \mathbb{N}^{d_1^{\prime} \times d_2^{\prime} \times d_3}$, $\bm{A}(\cdot): \mathbb{N}^{d_1 \times d_2} \rightarrow \mathbb{N}^{d_1^{\prime} \times d_2^{\prime}}$,  $\bm{s}_{{\bm{\theta}}^*}$, $\bm{s}_{{\bm{\phi}}^*}$, $\{\sigma_i\}_0^1$, $N$, $K$, $\lambda$
\\

\State $\bm{X}_N \sim \mathcal{N}(\bm{0}, \sigma^2_1\bm{I}) \in \mathbb{N}^{d_1 \times d_2 \times d_3}$
\For{$i$ \textbf{in} $N-1:0$}
\State $t \gets \frac{i}{N}$
\State $\bm{X}_i \gets$ torch.empty\_like($\bm{X}_N$)
\If{$\mod(i, K) \neq 0$}
\For{$j$ \textbf{in} $1:d_{3}$}
\State $\bm{x} \gets \bm{X}_{i+1}[:,:,j]$
\State $\bm{y} \gets \bm{Y}[:,:,j]$
\State $\Hat{\bm{x}}_0 \gets \bm{x} + \sigma_t^2 \cdot \bm{s}_{{\bm{\theta}}^*}(\bm{x}, t)$ 
\State $\bm{x}^{\prime} \gets$ step\_2D\_DPM($\bm{x}$, $\bm{s}_{{\bm{\theta}}^*}$, $\sigma_t$, $t$)
\State $\bm{x}^{\prime\prime} \gets \bm{x}^{\prime} - \lambda \nabla_{\bm{x}} \lVert \bm{A}(\Hat{\bm{x}}_0) - \bm{y} \rVert ^2 _2$
\State $\bm{X}_{i}[:,:,j] \gets \bm{x}^{\prime\prime}$
\EndFor
\Else
\For{$j$ \textbf{in} $1:d_{1}$}
\State $\bm{x} \gets \bm{X}_{i+1}[j,:,:]$
\State $\bm{x}^{\prime} \gets$ step\_2D\_DPM($\bm{x}$, $\bm{s}_{{\bm{\phi}}^*}$, $\sigma_t$, $t$)
\State $\bm{X}_i[j,:,:] \gets \bm{x}^{\prime}$
\EndFor
\EndIf
\EndFor
\State \Return $\bm{X}_0$
\end{algorithmic}
\end{algorithm}

\section{Methods}

In this paper, we investigate various applications of TPDM, which include medical domain 
inverse problems such as 1) MRI Z-axis (vertical axis) super-resolution (MR-ZSR), 2) compressed sensing MRI (CS-MRI), and 3) sparse view CT (SV-CT). 
%
%
In addition to solving the 3D inverse problem that applies conditioned sampling of the diffusion model, 4) TPDM is also used to generate unconditioned high-fidelity 3D voxels volumes in Brain MRI.

\subsection{Dataset}

The MR-ZSR, CS-MRI, and the 3D volume voxel generation task used our IRB-approved in-house brain MRI image dataset (\ie \verb|BMR-ZSR-1mm| and \verb|BMR-ZSR-5mm|). For detailed information, see Appendix~\ref{supp-bmr-zsr-dataset}. All volumes are in the shape of a 256$\times$256$\times$256 cube and standard 3T T1-weighted images. \verb|BMR-ZSR-1mm|, which is used for training and retrospective evaluation, has a 1mm slice thickness. 923 volumes (236,288 2D images) were used as a training dataset, and 1 volume was used as a test dataset with the retrospective slice thickness degradation or the CS-MRI subsampling simulation. \verb|BMR-ZSR-5mm|, which is a prospective dataset acquired at a slice thickness of 5mm, was used for the prospective clinical evaluation of MR-ZSR.

SV-CT task used the public CT dataset provided in the AAPM 2016 CT low-dose grand challenge~\cite{mccollough2017low}. The dataset consists of a total of 10 volumes of contrast-enhanced abdominal CT. To make the volume a $256\times256\times256$ cube, we resized the XY-plane to $256\times256$ and cropped the common part in the Z-direction to make the length 256 (\ie \verb|LDCT-CUBE| Dataset). One of the 10 volumes was used as the test dataset with the retrospective measurement simulation, and the remaining 9 were used as the training dataset. The data we used for training was only 2304 2D images so that we can demonstrate reliable performance even when the training data was small.

\subsection{Measurement model for inverse problems}
\noindent \textbf{MR-ZSR.} 
The goal of this task is to perform super-resolution of a 5mm slice thickness MRI image to 1mm slice resolution for quantitative brain MR analysis such as cortical thickness 
 measurement. Considering the slice selection process of MRI, the forward measurement kernel can be modeled by combining adjacent voxels in the Z-axis direction with averaging operation. For example, for 5mm to 1mm slice super-resolution, the forward model is an operation of degrading a 1mm slice image by grouping 5 adjacent XY-plane along the Z-axis direction and averaging each group to get a 5mm slice image. Here, we defined the number of pixels in the Z-axis direction of a group to be merged as \textit{merge size} ($M$).

We used the forward kernel just presented when creating the retrospective degraded MRI dataset (1mm $\rightarrow$ 2, 5mm).
We also used a slightly different forward measurement kernel used in the DPS step when solving the inverse problem MR-ZSR. The kernel is similar to the averaging process, but when averaging, divide by $\sqrt{M}$ instead of dividing by $M$, which is inspired by Song \etal~\cite{song2021scorebased} and Chung \etal~\cite{chung2022improving}'s diffusion model-based image colorization method.\\

\noindent \textbf{CS-MRI and SV-CT.} The forward measurement kernel for compressive sensing MRI (CS-MRI) involves applying a 2D subsampling mask to each slice of the image after transforming it into a k-space using a 2D Fourier transform. The resulting measurement $\bm{y}$ is given in the k-space domain.
In the case of sparse view CT (SV-CT), the forward measurement kernel is determined by the sparse view CT acquisition scenario, where angular projection views are subsampled at a sparse set of angles. The measurements are given in sinogram space.

\subsection{DPM training and sampling}
Both the MRI model and the CT model were trained and inferred under the common model setup and algorithms. The 2D image diffusion model constituting the TPDM used \verb|ncsnpp|~\cite{song2021scorebased} using VE-SDE which is scheduled by a geometric sequence $\sigma_0$=0.01 to $\sigma_1$=378. All inputs were normalized between 0 to 1. In the MR-ZSR problem, the YZ-plane (coronal) was used for the primary model and the XY-plane (axial) was used for the auxiliary model. In all other problems, the XY-plane (axial) was used for the primary model and the YZ plane (coronal) for the auxiliary model. The training was conducted with batch size 8, and the MRI model and CT model performed 300K and 100K training iterations, respectively. For the sampling stage, $N$=2000 and predictor-corrector sampling~\cite{song2021scorebased} method were employed.

\subsection{Comparison methods and evaluation}
For the 3D medical inverse problem, our method was compared with DiffusionMBIR~\cite{chung2022solving}, DPS~\cite{chung2023diffusion}, MCG~\cite{chung2022improving}, score-MRI~\cite{chung2022score}, score-CT~\cite{song2022solving}, L1-Wavelet~\cite{lustig2007sparse}, FBPConvNet~\cite{jin2017deep} and ADMM-TV. DiffusionMBIR is the state-of-the-art method to solve general 3D inverse problems which outperformed existing methods such as Score-MRI, DuDoRNet~\cite{zhou2020dudornet}, U-Net~\cite{ronneberger2015unet} and Zero-filled in CS-MRI and outperformed previous methods such as MCG, Lahiri~\etal~\cite{lahiri2023sparse}, FBPConvNet, and ADMM-TV in SV-CT. As the MR-ZSR problem is a new endeavor, no diffusion-based method has been devised to address it specifically. Quantitative evaluation was performed using peak-signal-to-noise-ratio (PSNR) and structural similarity index measure (SSIM)~\cite{wang2004image} for the retrospective test dataset. PSNR was evaluated in a 3D volume, and SSIM measured the average value of results of 2D slices for each slice direction (axial, coronal, and sagittal).

For the evaluation of MR-ZSR's clinical implications, seven patients with ischemic stroke were included in the evaluation (\verb|BMR-ZSR-5mm|). Visual assessments of cortical atrophy and white matter hyperintensity were conducted using the Global Cortical Atrophy scale~\cite{harper2015using} and Fazekas grade~\cite{scheltens1998white}, respectively. Using the TDPM, the prospective standard T1-weighted images with a 5 mm thickness were reconstructed into 1mm images. Five out of seven patients had 3D volumetric 1 mm T1-weighted images acquired simultaneously with a 5 mm T1-weighted image. The mean cortical thickness obtained with an upscaled T1-weighted image was compared to the mean cortical thickness measured with a 1mm raw T1-weighted image as the ground truth. Using FreeSurfer~\cite{tustison2014large} and ATROSCAN (JLK Inc., Seoul, Republic of Korea) based on Swin U-net~\cite{dahan2022surface}, the cortical thickness was measured.

\section{Experimental Results}

\subsection{MRI Z-axis super-resolution (MR-ZSR)} 

\begin{figure*}
\begin{center}
\includegraphics[width=\textwidth]{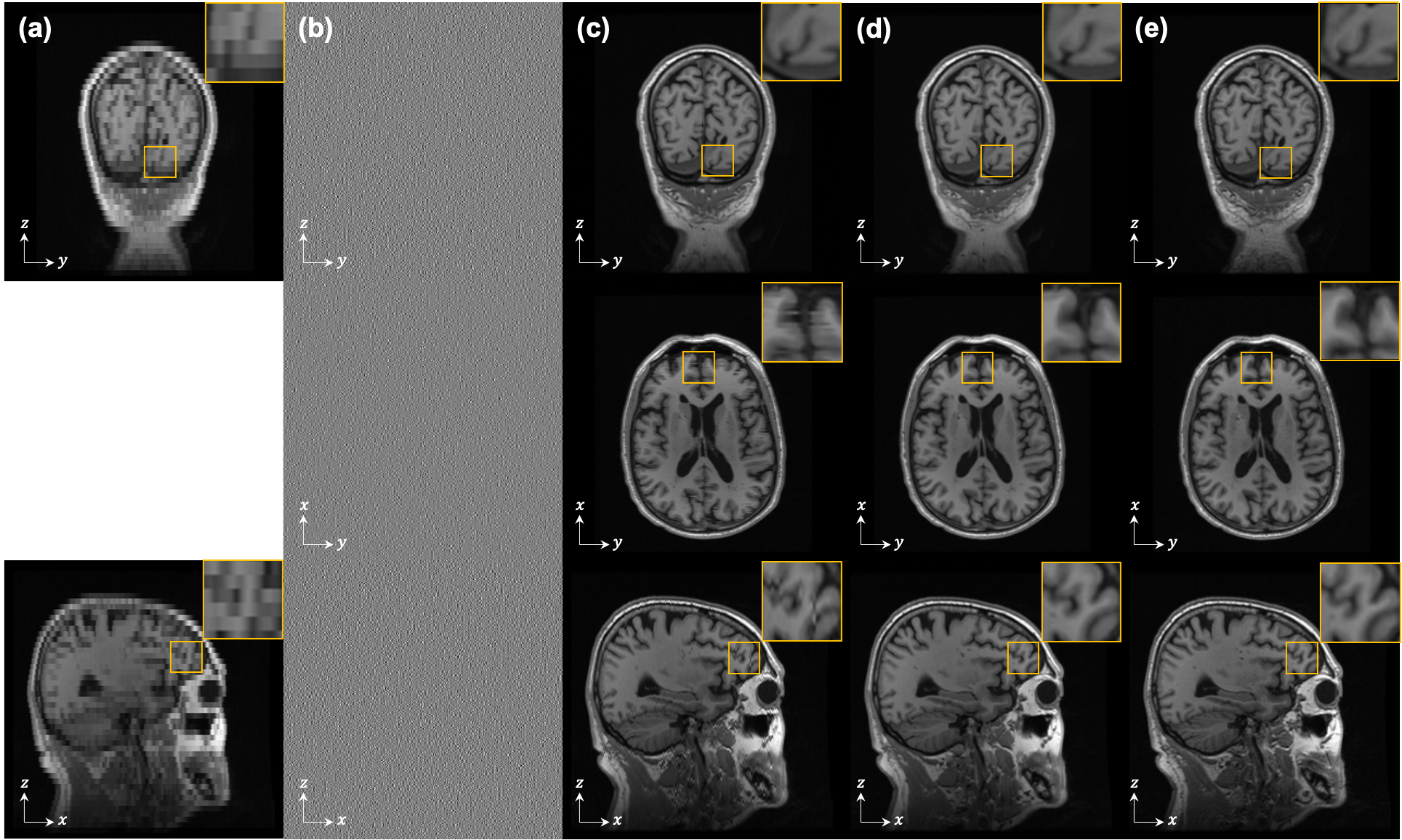}
\end{center}
\caption{MR-ZSR results from 5mm$\rightarrow$1mm ($\times$5)  of the retrospective test volume (first row: coronal slice, second row: axial slice, third row: sagittal slice). (a) measurement, (b) DiffusionMBIR~\cite{chung2022solving}, (c) DPS~\cite{chung2023diffusion}, (d) proposed method, (e) ground truth. For (d), first row: primary plane, second row: auxiliary plane.}
\label{fig:mrzsr_result_main}
\end{figure*}

\begin{table}
\begin{center}
\setlength{\tabcolsep}{0.45em}
\begin{tabular}{lllll}
\Xhline{3\arrayrulewidth}
                                      & \multirow{2}{*}{\textbf{PSNR} $\uparrow$} & \multicolumn{3}{c}{\textbf{SSIM $\uparrow$}}   \\ \cline{3-5}
\textbf{Method}                       &       & {\footnotesize Axial$^+$} & {\footnotesize Coronal$^{*}$} & {\footnotesize Sagittal} \\ \hline
TPDM (ours)                           & \textbf{35.97} & \textbf{0.970} & \textbf{0.966} & \textbf{0.964} \\
TPDM-MEAN                             & 32.84 & 0.963 & 0.957 & 0.955 \\
TPDM-MCG                       & 34.48 & 0.961 & 0.955 & 0.954 \\ \hline
DiffusionMBIR~\cite{chung2022solving} & \multicolumn{4}{c}{N/W} \\
DPS~\cite{chung2023diffusion}         & 34.77 & 0.965 & 0.963 & 0.960 \\
MCG~\cite{chung2022improving}         & 32.72 & 0.951 & 0.948 & 0.944 \\ \Xhline{3\arrayrulewidth}
\end{tabular}
\end{center}
\caption{Quantitative evaluation (PSNR, SSIM) of MR-ZSR (5mm$\rightarrow$1mm; $\times$5) on the BMR-ZSR-1mm test set. TPDM-MCG: TPDM uses MCG instead of DPS, TPDM-MEAN: The forward model used to create the retrospective dataset is used. N/W: Not Working. $^*$: primary plane, $^+$: auxiliary plane.}
\label{tab:mrzsr_result_main}
\end{table}

We first conducted MRI Z-axis $\times$5 super-resolution images of 5mm slices, which are mainly taken in clinical practice, to 1mm with the retrospective 5mm test dataset, and the results are in the Table~\ref{tab:mrzsr_result_main} and Fig.~\ref{fig:mrzsr_result_main}. For another merge size, see Appendix~\ref{mr-zsr-result-supp}. MR-ZSR using TPDM showed quantitatively better results than any other diffusion-based 2D/3D inverse problem-solving methods~\cite{chung2022solving, chung2023diffusion, chung2022improving}, and no artifacts occurred in any slice direction of the volume. In addition, the use of the auxiliary model not only improves the quality of the slice in the auxiliary direction but also has the effect of improving the detail of the entire slice directions (see (c) and (d) of the Fig.~\ref{fig:mrzsr_result_main}).

Notably, DiffusionMBIR~\cite{chung2022solving}, which is known to be the highest-performing general linear 3D inverse problem solver, did not work at all for our custom-designed MR-ZSR forward measurement kernel. This problem is caused by the total variation loss term, which is a key point loss term that gives consistency in the batch direction in DiffusionMBIR.

As a forward model of TPDM, when merging slices, dividing the sum of slices by $\sqrt{M}$ (TPDM) instead of using $N$ (TPDM-MEAN) yielded superior results. In addition, as a method for imposing measurement consistency constraints on the generation of the main model, TDPM with DPS (TPDM) exhibited superior outcomes than TDPM with MCG (TPDM-MCG), which is consistent with~\cite{chung2023diffusion}.

\begin{figure}
\begin{center}
\includegraphics[width=0.75\columnwidth]{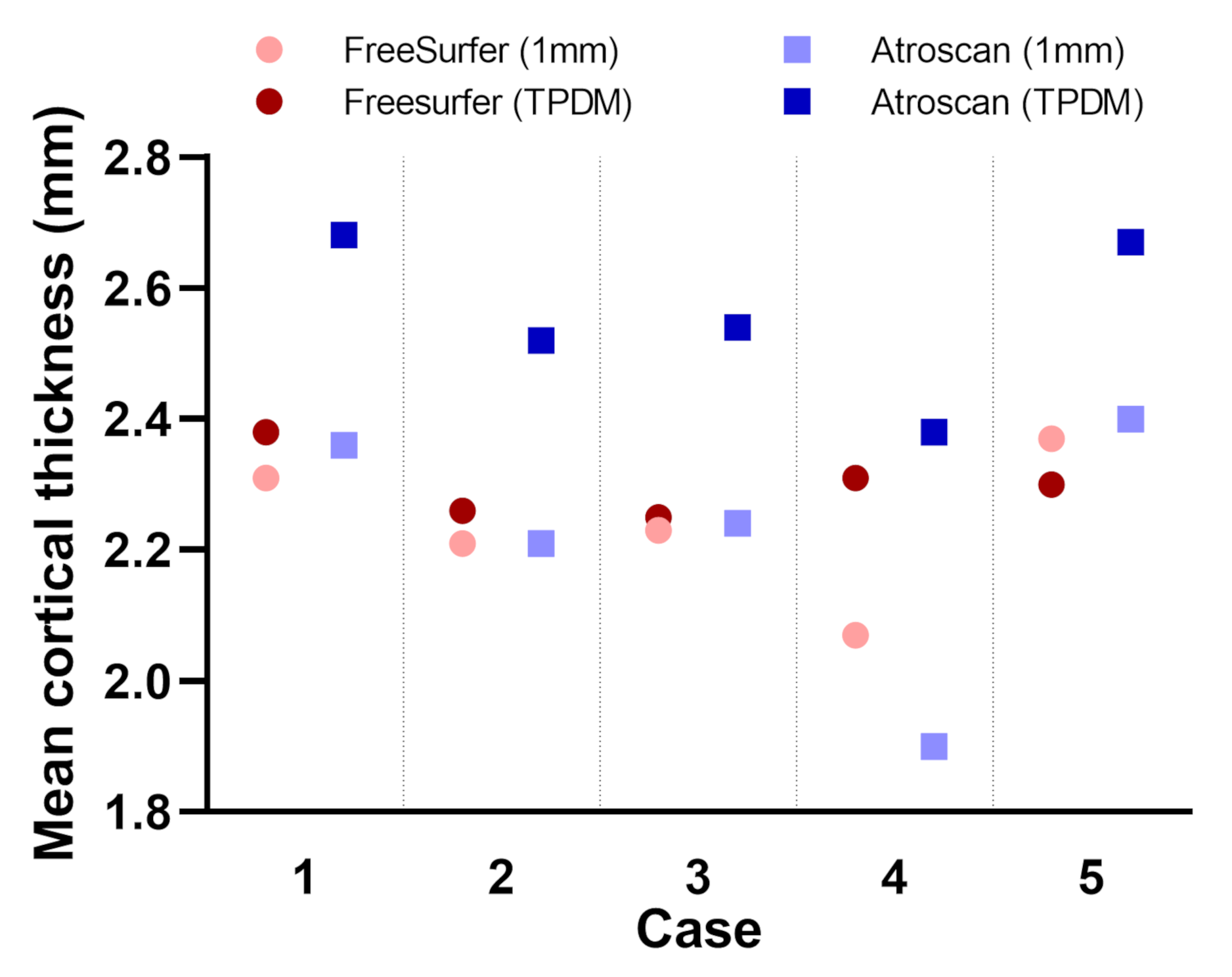}
\end{center}
\caption{Comparison of the brain cortical thickness measurement result of paired ground truth 1mm volumes (1mm) and upscaled volumes from 5mm to 1mm with TPDM MR-ZSR.}
\label{fig:mrzsr_pro_1_1}
\end{figure}

\begin{figure}
\begin{center}
\includegraphics[width=\columnwidth]{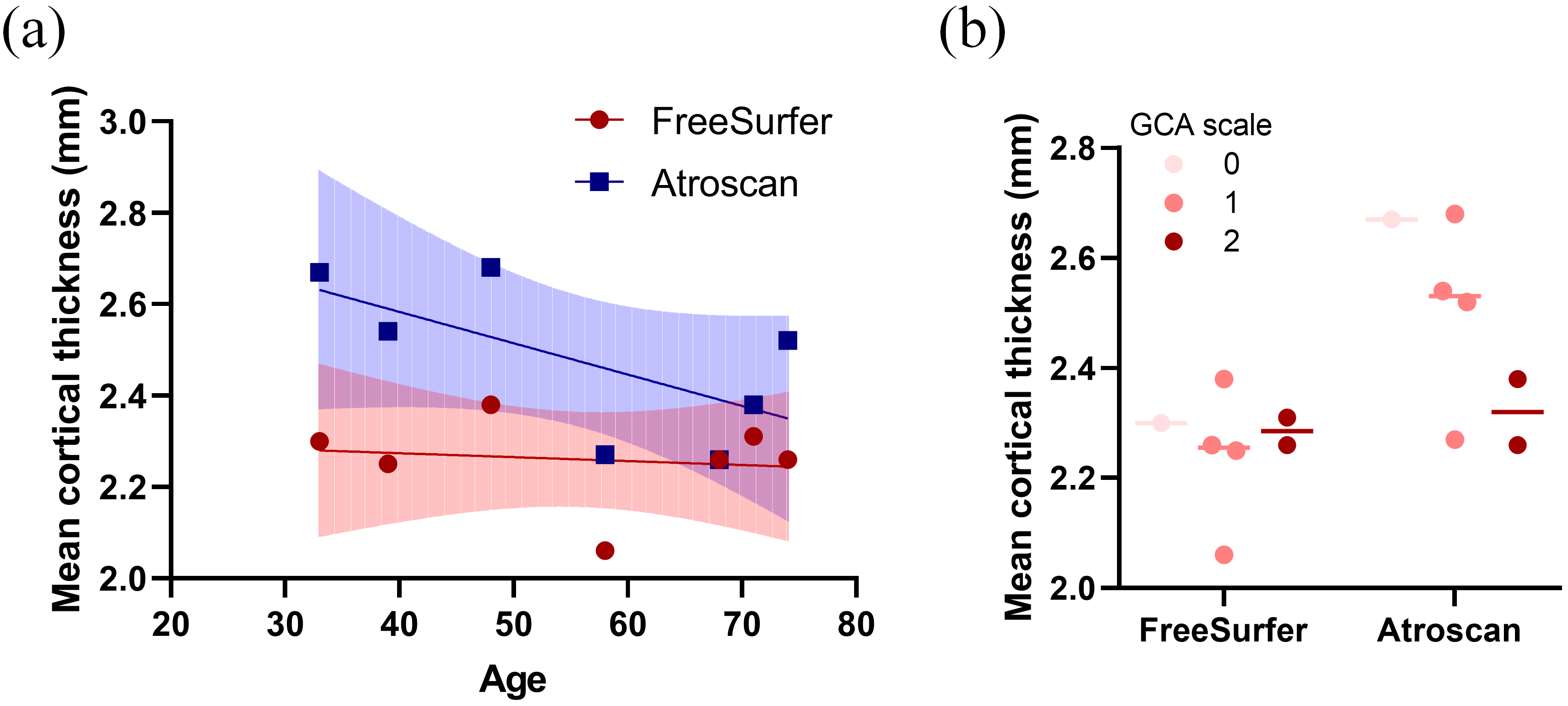}
\end{center}
\caption{Relationship between mean cortical thickness, and age (a) and GCA scale (b) measured through brain MRI volume that upscaled from 5mm to 1mm with TPDM MR-ZSR.}
\label{fig:mrzsr_pro_2}
\end{figure}

\begin{figure*}
\begin{center}
\includegraphics[width=\textwidth]{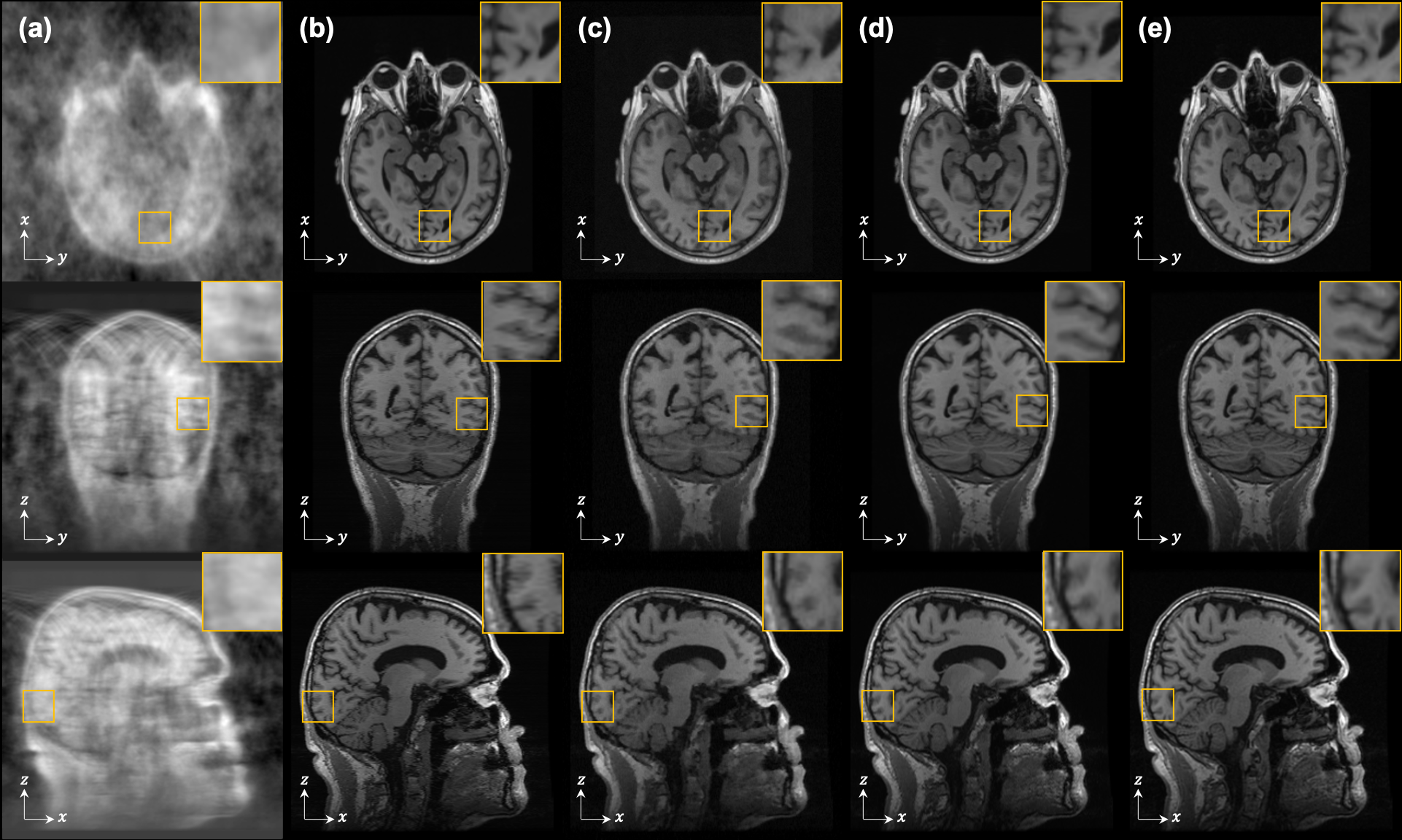}
\end{center}
   \caption{$\times48$ acceleration Poisson sub-sampled CS-MRI reconstruction results of the retrospective test volume (first row: axial slice, second row: coronal slice, third row: sagittal slice). (a) Measurement, (b) DPS~\cite{chung2023diffusion}, (c) DiffusionMBIR~\cite{chung2022solving}, (d) proposed method, (e) ground truth. For (d), first row: primary plane, second row: auxiliary plane.}
\label{fig:csmri_result_main}
\end{figure*}

The cortical mask measured in the reconstructed 1mm image was comparable to the mask estimated in the raw 1mm image by FreeSurfer (Appendix.~\ref{mr-zsr-result-supp}), with a mean difference of 0.06$\pm$0.11 (paired t-test, $p$=0.28), indicating that reconstructed T1 images by TDPM are reliably used for cortical thickness measurement that was not available in routine T1 image in clinical practice (Fig.~\ref{fig:mrzsr_pro_1_1}). When the cortical thickness was measured by ATROSCAN, the cortical mask in the reconstructed 1mm image was larger than the cortical mask in the raw 1mm image; the mean difference was 0.34$\pm$0.08 (paired t-test; $p$\textless0.001). Nonetheless, the difference is in a quite reasonable range for clinical uses. 

Although age-dependent cortical thickness decline was not clear in Freesurfer in the reconstructed T1 images, its general
tendency was clearly observed in ATROSCAN as demonstrated by the dot plot (Fig.~\ref{fig:mrzsr_pro_2}A). A similar trend was observed in the Global Cortical Atrophy scale (GCA), where cortical thickness by ATROSCAN was better correlated with than that by FreeSurfer (Fig.~\ref{fig:mrzsr_pro_2}B), albeit the difference between scales was not significant due to the small sample ($p$=0.82 and 0.22, respectively). 

 Routine brain MR T1 images are typically acquired with 5mm thickness to save scan time. The findings from this study suggest that the TDPM model could significantly expand the pool of eligible images for volumetric measurement, which would facilitate cognitive decline research. This is particularly important given that current routine 5mm acquisition protocols are inadequate for such research. Further investigations with larger sample sizes and more diverse populations will be needed to fully demonstrate the clinical implications of image reconstruction with TDPM.

\subsection{Compressed-sensing MRI (CS-MRI)}

\begin{table}
\begin{center}
\setlength{\tabcolsep}{0.45em}
\begin{tabular}{lllll}
\Xhline{3\arrayrulewidth}
                                      & \multirow{2}{*}{\textbf{PSNR} $\uparrow$} & \multicolumn{3}{c}{\textbf{SSIM $\uparrow$}}   \\ \cline{3-5}
\textbf{Method}                       &       & {\footnotesize Axial$^{*}$} & {\footnotesize Coronal$^+$} & {\footnotesize Sagittal} \\ \hline
TPDM (ours)                           & \textbf{37.17} & \textbf{0.966} & \textbf{0.967} & \textbf{0.965} \\ \hline
DiffusionMBIR~\cite{chung2022solving} & 34.83 & 0.907 & 0.909 & 0.906 \\
ADMM-TV                               & 27.01 & 0.812 & 0.802 & 0.812 \\ \hline
DPS~\cite{chung2023diffusion}         & 35.30 & 0.950 & 0.951 & 0.949 \\
score-MRI~\cite{chung2022score}       & 32.75 & 0.849 & 0.853 & 0.855 \\
L1-Wavelet~\cite{lustig2007sparse}    & 23.15 & 0.557 & 0.530 & 0.535 \\ \Xhline{3\arrayrulewidth}
\end{tabular}
\end{center}
\caption{Quantitative evaluation (PSNR, SSIM) of CS-MRI (Poisson, $\times$48 acc.) on the BMR-ZSR-1mm test set. $^*$: primary plane, $^+$: auxiliary plane.}
\label{tab:csmri_result_main}
\end{table}

We also evaluated TPDM by performing reconstruction on retrospective $\times$48 acceleration Poisson sub-sampled CS-MRI volumes (Fig.~\ref{fig:csmri_result_main}, Table~\ref{tab:csmri_result_main}). For other acceleration factors, see Appendix~\ref{cs-mri-result-supp}. Similarly to the outcomes by MR-ZSR, TPDM showed the best results compared to the prior art 2D/3D reverse problem-solving methods. Fig.~\ref{fig:csmri_result_main} also shows that TPDM accurately reconstructed the details, surpassing all other methods.

\subsection{Sparse-view CT (SV-CT)} \label{result-svct}


\begin{figure*}
\begin{center}
\includegraphics[width=\textwidth]{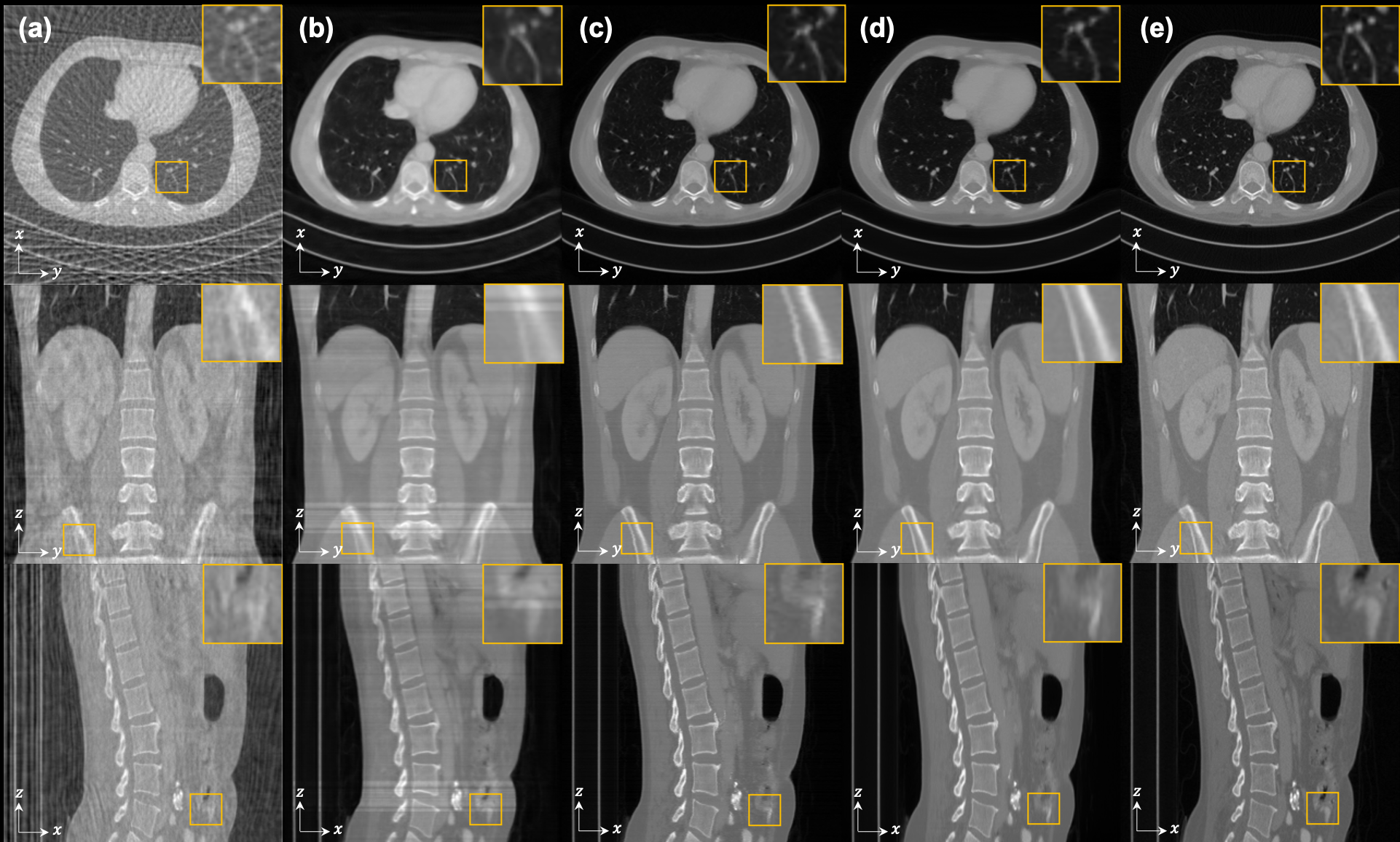}
\end{center}
\caption{36-view SV-CT reconstruction results of the retrospective test volume (first row: axial slice, second row: coronal slice, third row: sagittal slice). (a) Measurement, (b) FBPConvNet~\cite{jin2017deep}, (c) DPS~\cite{chung2023diffusion}, (d) proposed method, (e) ground truth. For (d), first row: primary plane, second row: auxiliary plane.}
\label{fig:svct_result_main}
\end{figure*}

\begin{table}
\begin{center}
\setlength{\tabcolsep}{0.45em}
\begin{tabular}{lllll}
\Xhline{3\arrayrulewidth}
                                      & \multirow{2}{*}{\textbf{PSNR} $\uparrow$} & \multicolumn{3}{c}{\textbf{SSIM $\uparrow$}}   \\ \cline{3-5}
\textbf{Method}                       &       & {\footnotesize Axial$^{*}$} & {\footnotesize Coronal$^+$} & {\footnotesize Sagittal} \\ \hline
TPDM (ours)                           & \textbf{38.25} & \textbf{0.947} & \textbf{0.951} & \textbf{0.949} \\ \hline
DiffusionMBIR~\cite{chung2022solving} & 34.78 & 0.857 & 0.856 & 0.861 \\
ADMM-TV                               & 30.33 & 0.856 & 0.894 & 0.867 \\  \hline
DPS~\cite{chung2023diffusion}         & 38.20 & 0.942 & 0.943 & 0.941 \\
score-CT~\cite{song2022solving}       & 37.56 & 0.922 & 0.922 & 0.924 \\
FBPConvNet~\cite{jin2017deep}         & 32.09 & 0.945 & 0.932 & 0.931 \\  \Xhline{3\arrayrulewidth}
\end{tabular}
\end{center}
\caption{Quantitative evaluation (PSNR, SSIM) of SV-CT (36-view) on the LDCT-CUBE test set. $^*$: primary plane, $^+$: auxiliary plane. Note that only 2304 2D images were used for training.}
\label{tab:svct_result_main}
\end{table}

The CT problem was used for only 9 volumes (about 2000 2D images) as a train dataset data to test the performance of TPDM in extremely small data conditions. The experimental results for 36-view SV-CT are shown in Table~\ref{tab:svct_result_main}, Fig.~\ref{fig:svct_result_main}. Despite training with a highly limited dataset, the TPDM model performed well compared to the other models. Although the quantitative improvement over DPS~\cite{chung2023diffusion} is not large, TPDM outperforms DPS significantly due to DPS being a 2D inverse problem solver which introduces artifacts in the batch direction when applied to 3D inverse problems. In the case of FBPConvNet~\cite{jin2017deep}, the SSIM in the axial direction has a small improvement, but since it is also a 2D model, it exhibits poor performance for other slice directions. Furthermore, the blurred outcomes commonly observed in convolutional networks trained using supervised techniques remain evident.

\subsection{Unconditional 3D voxels volume generation}

\begin{figure*}
\begin{center}
\includegraphics[width=1.25\columnwidth]{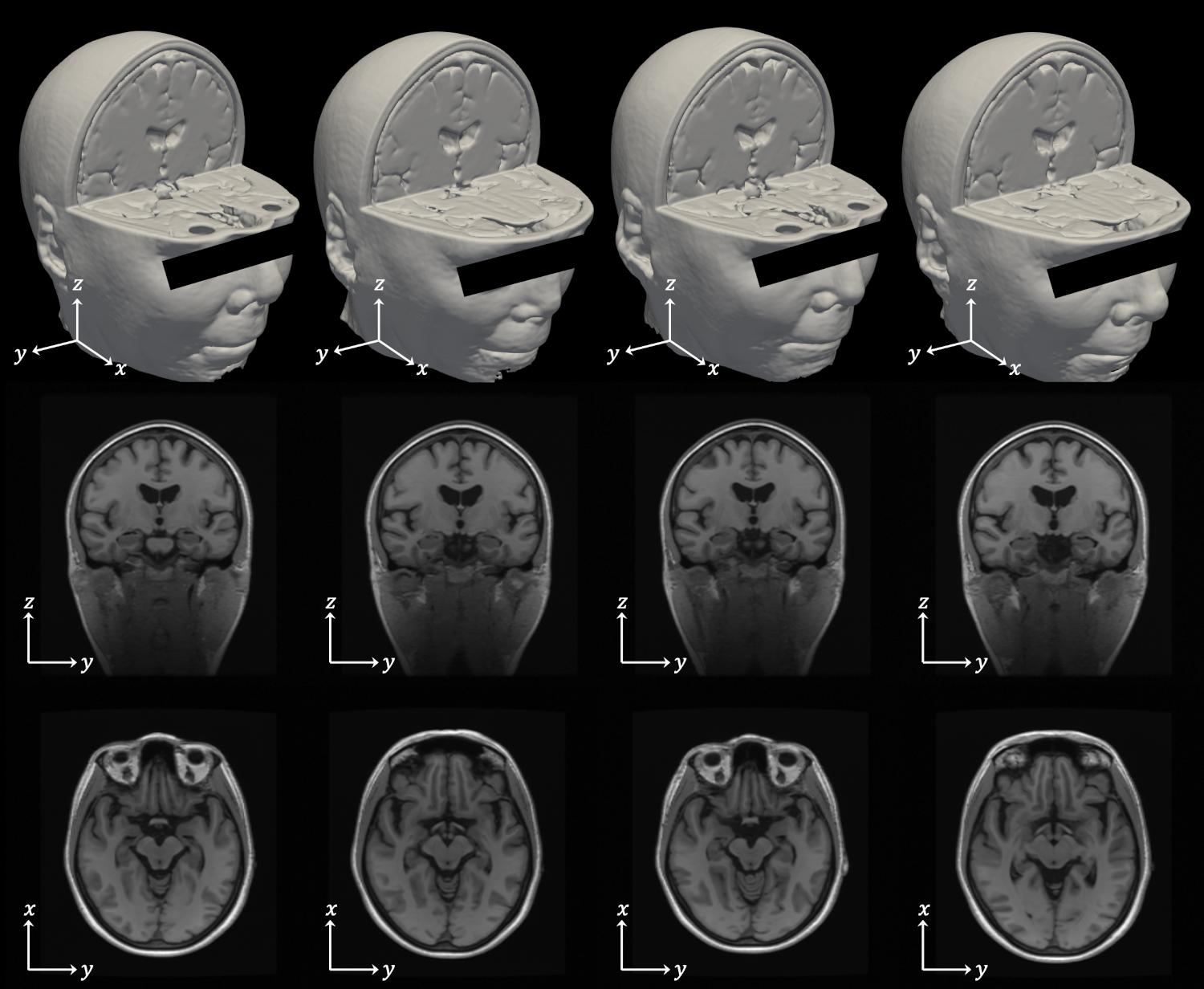}
\end{center}
   \caption{Results of the human head MRI volume generation using unconditioned TPDM. To visualize the volume, the iso-surface contour is expressed as a surface after removing a quarter of the volume.}
\label{fig:uncond_result_main}
\end{figure*}

Using TPDM, we attempted to generate a full 3D voxel volume unconditionally (for the algorithm, see Appendix~\ref{tpdm-uncond}). We trained the TPDM model using the \verb|BMR-ZSR-1mm| dataset and used it to generate an MRI volume of the human head, the results of which are presented in Fig.~\ref{fig:uncond_result_main}. Notably, we were able to create a complete three-dimensional voxel volume with high resolution and quality, without relying on any measurement guidance. We believe that TPDM's ability to generate 3D volumes is not solely due to the 2D image order guidance provided by the measurement in the DPS step, but also due to the alternative denoising algorithms of the two diffusion models. The empirical evidence presented here supports the reasonableness of our proposed data distribution assumptions.

\section{Conclusion}

In this study, we introduced TPDM, a method for solving the general 3D inverse problem and generating voxels volume with pre-trained two perpendicular 2D diffusion models. TPDM works in a completely unsupervised manner and does not require any fine-tuning for individual inverse problems. It handles 3D volume without using any 3D diffusion model by assuming a 3D distribution as the product distribution of 2D distributions, effectively avoiding the curse of dimensionality while still utilizing probability distributions of 3D volume. Our findings indicated that TPDM outperforms existing state-of-the-art 3D inverse problem-solving methods on several medical 3D reconstruction problems, even when trained with a significantly limited amount of data. Finally, using TPDM and a novel forward measurement model, we first-ever attempted diffusion-based Z-directional super-resolution of MRI images and demonstrated exceptional outcomes in both technical and clinical aspects.

\section*{Acknowledgement}
This work was supported by the National Research Foundation of Korea under Grant NRF-2020R1A2B5B03001980, and by Field-oriented Technology Development Project for Customs Administration through National Research Foundation of Korea(NRF) funded by the Ministry of Science \& ICT and Korea Customs Service(NRF-2021M3I1A1097938).

\clearpage
\clearpage
{\small
\bibliographystyle{ieee_fullname}
\bibliography{egbib}
}

\clearpage
\appendix

\section*{Supplementary Material}

\section{Additional algorithms}
\subsection{Training TPDM} \label{training-tpdm}
TPDM can be trained with a three-dimensional volume dataset and Algorithm~\ref{alg:train_tpdm}.

\begin{algorithm}
\caption{Training TPDM}\label{alg:train_tpdm}
\begin{algorithmic}
\Require $\{\bm{X}_i \in \mathbb{N}^{d_1 \times d_2 \times d_3} \}_1^M$, $\{\sigma_i\}_0^1$

\State $D_{prim} \gets \{\}$, $D_{aux} \gets \{\}$ \Comment{Create 2D datasets}
\For{$i$ \textbf{in} $1:M$}
\For{$j$ \textbf{in} $1:d_{3}$}
\State $D_{prim}$.add($\bm{X}_i[:, :, j]$)
\EndFor
\For{$j$ \textbf{in} $1:d_{1}$}
\State $D_{aux}$.add($\bm{X}_i[j, :, :]$)
\EndFor
\EndFor

\State $\bm{s}_{{\bm{\theta}}^*} \gets $ train\_2D\_DPM($D_{prim}$, $\{\sigma_i\}_0^1$) \Comment{Train DPMs}
\State $\bm{s}_{{\bm{\phi}}^*} \gets $ train\_2D\_DPM($D_{aux}$, $\{\sigma_i\}_0^1$)

\State \Return $\bm{s}_{{\bm{\theta}}^*}$, $\bm{s}_{{\bm{\phi}}^*}$
\end{algorithmic}
\end{algorithm}

\subsection{Sampling with real value $K$} \label{real-value-k}
When $K$ is a real number, select the primary model and the auxiliary model in a stochastic way through sampling from the Bernoulli distribution with $p=1-1/K$ (Algorithm~\ref{alg:inference_tpdm_real_k}).

\begin{algorithm}
\caption{Solving 3D Inverse Problem with TPDM}\label{alg:inference_tpdm_real_k}
\begin{algorithmic}
\Require $\bm{Y} \in \mathbb{N}^{d_1^{\prime} \times d_2^{\prime} \times d_3}$, $\bm{A}(\cdot): \mathbb{N}^{d_1 \times d_2} \rightarrow \mathbb{N}^{d_1^{\prime} \times d_2^{\prime}}$,  $\bm{s}_{{\bm{\theta}}^*}$, $\bm{s}_{{\bm{\phi}}^*}$, $\{\sigma_i\}_0^1$, $N$, $K$, $\lambda$
\\

\State $\bm{X}_N \sim \mathcal{N}(\bm{0}, \sigma^2_1\bm{I}) \in \mathbb{N}^{d_1 \times d_2 \times d_3}$
\For{$i$ \textbf{in} $N-1:0$}
\State is\_primary $\sim  Bernoulli(1-\frac{1}{K})$
\State $t \gets \frac{i}{N}$
\State $\bm{X}_i \gets$ torch.empty\_like($\bm{X}_N$)
\If{is\_primary}
\For{$j$ \textbf{in} $1:d_{3}$}
\State $\bm{x} \gets \bm{X}_{i+1}[:,:,j]$
\State $\bm{y} \gets \bm{Y}[:,:,j]$
\State $\Hat{\bm{x}}_0 \gets \bm{x} + \sigma_t^2 \cdot \bm{s}_{{\bm{\theta}}^*}(\bm{x}, t)$ 
\State $\bm{x}^{\prime} \gets$ step\_2D\_DPM($\bm{x}$, $\bm{s}_{{\bm{\theta}}^*}$, $\sigma_t$, $t$)
\State $\bm{x}^{\prime\prime} \gets \bm{x}^{\prime} - \lambda \nabla_{\bm{x}} \lVert \bm{A}(\Hat{\bm{x}}_0) - \bm{y} \rVert ^2 _2$
\State $\bm{X}_{i}[:,:,j] \gets \bm{x}^{\prime\prime}$
\EndFor
\Else
\For{$j$ \textbf{in} $1:d_{1}$}
\State $\bm{x} \gets \bm{X}_{i+1}[j,:,:]$
\State $\bm{x}^{\prime} \gets$ step\_2D\_DPM($\bm{x}$, $\bm{s}_{{\bm{\phi}}^*}$, $\sigma_t$, $t$)
\State $\bm{X}_i[j,:,:] \gets \bm{x}^{\prime}$
\EndFor
\EndIf
\EndFor
\State \Return $\bm{X}_0$
\end{algorithmic}
\end{algorithm}

\subsection{3D voxel volume generation with TPDM} \label{tpdm-uncond}
TPDM's unconditional sampling can be performed by removing the DPS step of the primary model from the conditional sampling algorithm of TPDM (Algorithm~\ref{alg:uncond_tpdm}).

\begin{algorithm}
\caption{Unconditional Sampling with TPDM}\label{alg:uncond_tpdm}
\begin{algorithmic}
\Require $\bm{s}_{{\bm{\theta}}^*}$, $\bm{s}_{{\bm{\phi}}^*}$, $\{\sigma_i\}_0^1$, $N$, $K$

\State $\bm{X}_N \sim \mathcal{N}(\bm{0}, \sigma^2_1\bm{I}) \in \mathbb{N}^{d_1 \times d_2 \times d_3}$
\For{$i$ \textbf{in} $N-1:0$}
\State $t \gets \frac{i}{N}$
\State $\bm{X}_i \gets$ torch.empty\_like($\bm{X}_N$)
\If{$\mod(i, K) \neq 0$}
\For{$j$ \textbf{in} $1:d_{3}$}
\State $\bm{x} \gets \bm{X}_{i+1}[:,:,j]$
\State $\bm{x}^{\prime} \gets$ step\_2D\_DPM($\bm{x}$, $\bm{s}_{{\bm{\theta}}^*}$, $\sigma_t$, $t$)
\State $\bm{X}_{i}[:,:,j] \gets \bm{x}^{\prime}$
\EndFor
\Else
\For{$j$ \textbf{in} $1:d_{1}$}
\State $\bm{x} \gets \bm{X}_{i+1}[j,:,:]$
\State $\bm{x}^{\prime} \gets$ step\_2D\_DPM($\bm{x}$, $\bm{s}_{{\bm{\phi}}^*}$, $\sigma_t$, $t$)
\State $\bm{X}_i[j,:,:] \gets \bm{x}^{\prime}$
\EndFor
\EndIf
\EndFor
\State \Return $\bm{X}_0$
\end{algorithmic}
\end{algorithm}

\section{Dataset}
\subsection{BMR-ZSR-5mm} \label{supp-bmr-zsr-dataset}
We generated a 1mm volumetric dataset (\ie \verb|BMR-ZSR-1mm|) using structural brain 3T T1-weighted images from the Alzheimer's Disease Neuroimaging Initiative (ADNI) dataset (271 subjects with probable dementia and 211 subjects with normal cognition) and data from a university hospital's voluntary health screening program (441 normal). Evaluation was performed on 1 subject from the ADNI dataset which has normal cognition with the retrospective slice thickness degradation or the CS-MRI sub-sampling simulation.

The prospective 5mm volumetric dataset (\ie \verb|BMR-ZSR-5mm|) is also structural brain 3T T1-weighted images, which is composed of seven patients with ischemic stroke. The clinical information of the subjects is in Table~\ref{tab:mrzsr_pro_subject_info}. Five out of seven patients had 3D volumetric 1mm T1-weighted images acquired simultaneously with a 5mm T1-weighted image.

\begin{table*}
\begin{center}
\setlength{\tabcolsep}{0.1em}
\begin{tabular}{|c|c|c|c|c|c|c|c|c|c|c|}
\Xhline{3\arrayrulewidth}
\# & Age & Gender & Cortical infarct & GCA scale & WMH grade & Previous stroke & Hypertension & Diabetes & Hyperlipidemia & Current smoking \\ \hline
1 & 74  & M      & No               & 1         & 0         & No              & Yes          & No       & No             & No              \\
2 & 71  & M      & Yes              & 2         & 2         & No              & Yes          & Yes      & No             & Yes             \\
3 & 33  & M      & No               & 0         & 0         & No              & No           & No       & No             & No              \\
4 & 48  & F      & No               & 1         & 2         & Yes             & No           & Yes      & No             & No              \\
5 & 39  & M      & Yes              & 1         & 1         & No              & No           & No       & No             & No              \\
6 & 58  & F      & No               & 1         & 2         & Yes             & No           & Yes      & Yes            & No              \\
7 & 68  & M      & No               & 2         & 1         & Yes             & Yes          & No       & No             & Yes            \\ \Xhline{3\arrayrulewidth}
\end{tabular}
\end{center}
\caption{Subject information of the prospective clinical evaluation of MR-ZSR.}
\label{tab:mrzsr_pro_subject_info}
\end{table*}

\subsection{LDCT-CUBE}
The \verb|LDCT-CUBE| dataset was built based on the contrast-enhanced abdominal CT presented in the AAPM 2016 CT low-dose grand challenge~\cite{mccollough2017low}. The data set was converted into 10 volumes with 256$\times$256 slices in the axial slice direction through the same method as in~\cite{chung2022solving} (\verb|LDCT|). Since \verb|LDCT| has different lengths in the vertical axis direction, a common part of volumes was manually selected and 256 consecutive slices were cropped to generate 256$\times$256$\times$256 cube-shaped volumes. Zero padding was added if the original slice was less than 256 slices. See Table~\ref{tab:aapm_crop_info} for the detailed cropping parameters.

\begin{table}
\begin{center}
\setlength{\tabcolsep}{0.75em}
\begin{tabular}{ccc}  \Xhline{3\arrayrulewidth}
Patient ID  & \# of raw slices & Cropped slices range \\  \hline
L096        & 658          & 224:480             \\
L109        & 254          & 000:254               \\
L143        & 468          & 212:468             \\
L192        & 480          & 064:320              \\
L286        & 420          & 000:256               \\
L291        & 685          & 249:505             \\
L310        & 426          & 030:286              \\
L333        & 488          & 049:305              \\
L506        & 421          & 000:256               \\ \hline
L067 (test) & 448          & 004:260               \\ \Xhline{3\arrayrulewidth}
\end{tabular}
\end{center}
\caption{Cropping information of the LDCT-CUBE dataset. The range shown includes the start point and does not include the endpoint. The indexes start at 0.}
\label{tab:aapm_crop_info}
\end{table}

\section{Additonal results}
\subsection{MRI Z-axis super-resolution (MR-ZSR)} \label{mr-zsr-result-supp}

\begin{figure}
\begin{center}
\includegraphics[width=\columnwidth]{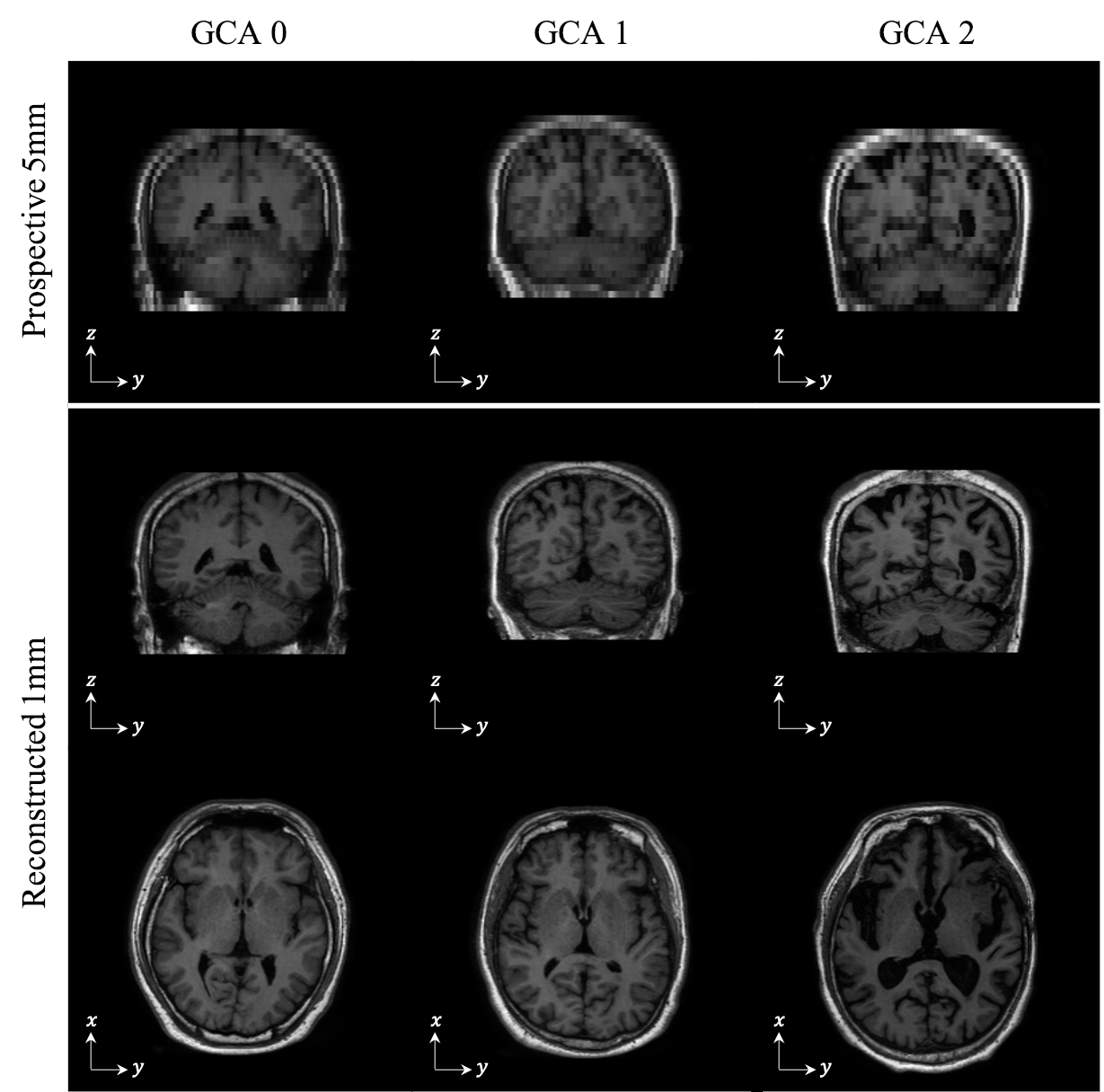}
\end{center}
\caption{Result of the prospective 5mm$\rightarrow$1mm ($\times$5) MR-ZSR for different GCA scales. first/second row: primary plane, third row: auxiliary plane.}
\label{fig:mrzsr_pro_gca}
\end{figure}

\begin{figure}
\begin{center}
\includegraphics[width=\columnwidth]{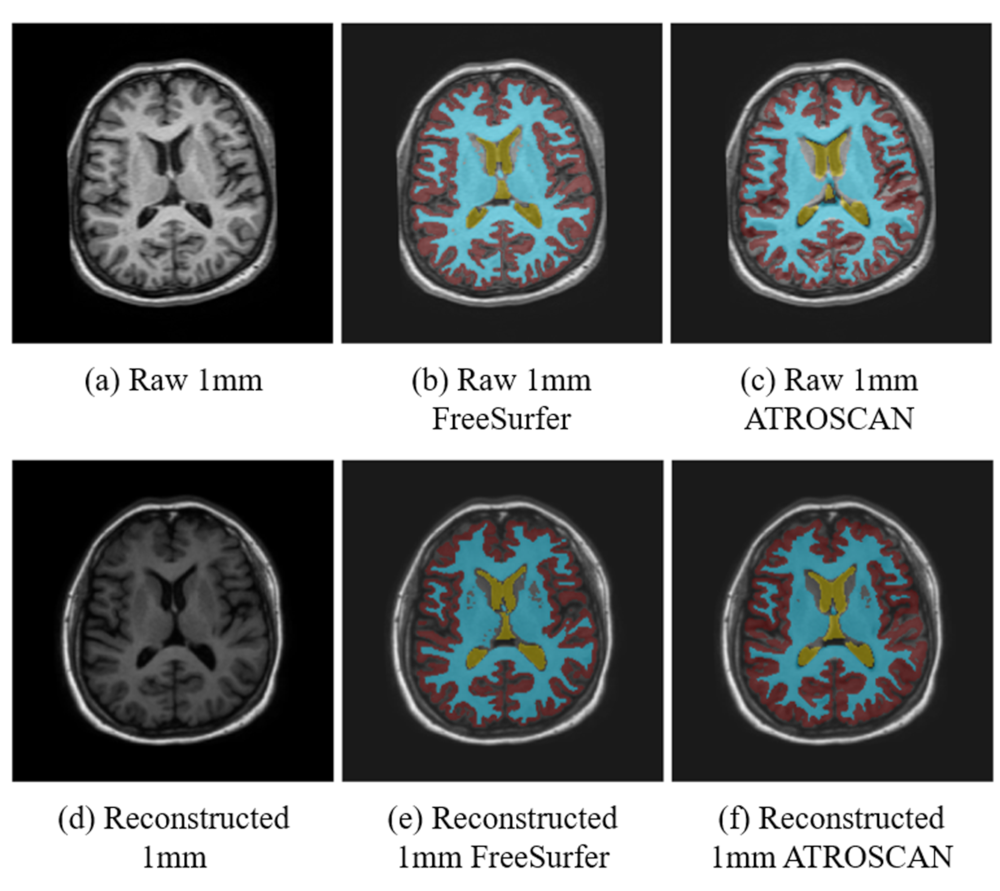}
\end{center}
\caption{Comparison of estimated cortical mask between raw 1mm image and 5mm$\rightarrow$1mm ($\times$5) image from the prospective test volume.}
\label{fig:mrzsr_pro_1_2}
\end{figure}

Additional results of the prospective clinical evaluation of a slice thickness of 5mm to 1mm MR-ZSR are shown in Fig.~\ref{fig:mrzsr_pro_gca}, Fig.~\ref{fig:mrzsr_pro_1_2}, and Table~\ref{tab:mrzsr_pro_cortical_result}. It was shown that MR-ZSR using TPDM works well for various GCA scales, especially in the presence of lesions. Although the \verb|BMR-ZSR-5mm| had slightly different MRI sequence parameters from the \verb|BMR-ZSR-1mm| used for training, TPDM was well adapted without any additional model modification. These reconstructed 1 mm images were evaluated as suitable for use as an input for a conventional cortical mask segmentation algorithm designed to operate only on images acquired with actual 1mm slice thickness.

Table~\ref{tab:mrzsr_supp_result} shows the results of the retrospective quantitative evaluation of MR-ZSR with a slice thickness of 3mm to 1mm.

\begin{table}
\begin{center}
\begin{tabular}{ccccc}
\Xhline{3\arrayrulewidth}
\multirow{3}{*}{\#} & \multicolumn{4}{c}{\textbf{Mean cortical thickness}}                       \\ \cline{2-5} 
                   & \multicolumn{2}{c}{Raw 1mm} & \multicolumn{2}{c}{TPDM 5mm $\rightarrow$ 1mm} \\ \cline{2-5} 
                   & FreeSurfer   & Astroscan    & FreeSurfer       & Astroscan        \\ \hline
1                  & 2.21±0.92  & 2.21±0.78  & 2.26±0.98        & 2.52±0.84      \\
2                  & 2.07±0.93  & 1.90±0.72  & 2.31±1.12        & 2.38±0.84      \\
3                  & 2.37±1.00  & 2.40±0.81  & 2.30±1.06        & 2.67±0.85      \\
4                  & 2.31±1.00  & 2.36±0.82  & 2.38±1.06        & 2.68±0.86      \\
5                  & 2.23±0.95  & 2.24±0.79  & 2.25±1.02        & 2.54±0.84      \\
6                  & N/A           & N/A           & 2.06±1.00        & 2.27±0.87      \\
7                  & N/A           & N/A           & 2.26±1.06        & 2.40±0.86      \\ \Xhline{3\arrayrulewidth}
\end{tabular}
\end{center}
\caption{Result of the mean cortical thickness measurement of prospective ground truth 1mm MRI volume and upscaled 1mm MRI volume from 5mm by TPDM.}
\label{tab:mrzsr_pro_cortical_result}
\end{table}

\begin{table}
\begin{center}
\setlength{\tabcolsep}{0.45em}
\begin{tabular}{lllll}
\Xhline{3\arrayrulewidth}
                                      & \multirow{2}{*}{\textbf{PSNR} $\uparrow$} & \multicolumn{3}{c}{\textbf{SSIM $\uparrow$}}   \\ \cline{3-5}
\textbf{Method}                       &       & {\footnotesize Axial$^+$} & {\footnotesize Coronal$^{*}$} & {\footnotesize Sagittal} \\ \hline
TPDM (ours)                           & \textbf{38.76} & \textbf{0.982} & \textbf{0.979} & \textbf{0.978} \\ \hline
DiffusionMBIR~\cite{chung2022solving} & \multicolumn{4}{c}{N/W} \\ \Xhline{3\arrayrulewidth}
\end{tabular}
\end{center}
\caption{Quantitative evaluation (PSNR, SSIM) of MR-ZSR (3mm$\rightarrow$1mm; $\times$3) on the BMR-ZSR-1mm test set. N/W: Not Working. $^*$: primary plane, $^+$: auxiliary plane.}
\label{tab:mrzsr_supp_result}
\end{table}

\subsection{Compressed-sensing MRI (CS-MRI)} \label{cs-mri-result-supp}

We further attempted reconstruction on $\times$8 and $\times$24 accelerated Poisson sub-sampled CS-MRI volumes. The results are Table~\ref{tab:csmri_supp_result_acc8} and Table~\ref{tab:csmri_supp_result_acc24}, respectively. If the problem is straightforward ($\times$8 acceleration), each 2D image can be restored with a high degree of accuracy, leading to near-perfect outcomes even with only the 2D solving method (DPS~\cite{chung2023diffusion}). Nevertheless, as the complexity of the challenge increases ($\times$24, $\times$48), we can only get better results in 3D with the assistance of a 3D prior.

\begin{table}
\begin{center}
\setlength{\tabcolsep}{0.45em}
\begin{tabular}{lllll}
\Xhline{3\arrayrulewidth}
                                      & \multirow{2}{*}{\textbf{PSNR} $\uparrow$} & \multicolumn{3}{c}{\textbf{SSIM $\uparrow$}}   \\ \cline{3-5}
\textbf{Method}                       &       & {\footnotesize Axial$^{*}$} & {\footnotesize Coronal$^+$} & {\footnotesize Sagittal} \\ \hline
TPDM (ours)                           & 44.96 & 0.988 & 0.989 & 0.988 \\ \hline
DiffusionMBIR~\cite{chung2022solving} & 41.21 & 0.934 & 0.934 & 0.934 \\
DPS~\cite{chung2023diffusion}         & \textbf{47.10} & \textbf{0.991} & \textbf{0.991} & \textbf{0.991} \\
score-MRI~\cite{chung2022score} & 39.90 & 0.914 & 0.914 & 0.913 \\ \Xhline{3\arrayrulewidth}
\end{tabular}
\end{center}
\caption{Quantitative evaluation (PSNR, SSIM) of CS-MRI (Poisson, $\times$8 acc) on the BMR-ZSR-1mm test set. $^*$: primary plane, $^+$: auxiliary plane.}
\label{tab:csmri_supp_result_acc8}
\end{table}

\begin{table}
\begin{center}
\setlength{\tabcolsep}{0.45em}
\begin{tabular}{lllll}
\Xhline{3\arrayrulewidth}
                                      & \multirow{2}{*}{\textbf{PSNR} $\uparrow$} & \multicolumn{3}{c}{\textbf{SSIM $\uparrow$}}   \\ \cline{3-5}
\textbf{Method}                       &       & {\footnotesize Axial$^{*}$} & {\footnotesize Coronal$^+$} & {\footnotesize Sagittal} \\ \hline
TPDM (ours)                           & \textbf{40.34} & \textbf{0.979} & \textbf{0.978} & \textbf{0.978} \\ \hline
DiffusionMBIR \cite{chung2022solving} & 37.48 & 0.895 & 0.899 & 0.897 \\
DPS \cite{chung2023diffusion}         & 39.06 & 0.965 & 0.967 & 0.965 \\ 
score-MRI~\cite{chung2022score} & 35.54 & 0.843 & 0.845 & 0.844 \\ \Xhline{3\arrayrulewidth}
\end{tabular}
\end{center}
\caption{Quantitative evaluation (PSNR, SSIM) of CS-MRI (Poisson, $\times$24 acc) on the BMR-ZSR-1mm test set. $^*$: primary plane, $^+$: auxiliary plane.}
\label{tab:csmri_supp_result_acc24}
\end{table}




\section{Sampling hyperparameters} \label{mr-zsr-sampling-supp}
Sampling hyperparameters utilized for TPDM, MCG~\cite{chung2022improving}, DPS~\cite{chung2023diffusion}, DiffusionMBIR~\cite{chung2022solving}, score-MRI~\cite{chung2022score}, and score-CT~\cite{song2022solving} are presented for each experiment. The sampling hyperparameters for all comparative experiments were configured to match the specific hyperparameters that yielded optimal results from the models identified during the experimental phase. Common to all experiments in TPDM, an integer value $K$=2 was used for the MRI model, and a real number value $K$=2.7 was used for the CT model. All diffusion models were sampled with N=2000 sampling steps, regardless of the problem.

\subsection{MR-ZSR}
Retrospective 5mm to 1mm used $\lambda$=4 for TPDM/DPS, $\lambda$=0.1 for MCG. Prospective 5mm to 1mm uses $\lambda$=1 for TPDM. Retrospective 3mm to 1mm was performed with TPDM by $\lambda$=2.

\subsection{CS-MRI}
For Poisson sub-sampled $\times$48 acceleration, $\lambda$=0.01 was used by TPDM/DPS. DiffusionMBIR used $\lambda$=0.0001 and $\rho$=0.1. For the $\times$24 acceleration, TPDM/DPS uses $\lambda$=0.007, and DiffusionMBIR uses $\lambda$=0.0001 and $\rho$=0.1. For $\times$8 acceleration, $\lambda$=0.002 for TPDM/DPS, and $\lambda$=0.0005 and $\rho$=0.1 for DiffusionMBIR. Score-MRI has no hyperparameters configuring the sampling stage.

\subsection{SV-CT}
For the 36-view SV-CT problem, $\lambda$=0.025 was used for TPDM/DPS, and $\lambda$=0.01 and $\rho$=40 were used for DiffusionMBIR. Score-CT used $\lambda$=0.8.

\section{Computational resources}
Both the training and sampling processes of the TPDM were executed utilizing two NVIDIA GeForce RTX 3090 GPUs. Employing the settings expounded upon in the text, the training duration for the MRI and CT models amounted to approximately 3 days and 1 day, respectively, for each 2D model, be it primary or auxiliary. The process of TPDM sampling necessitated an approximate timeframe of 24 to 36 hours per volume, contingent upon the specific problem type. Adopting a batch size of 6 during sampling, TPDM consumption of VRAM totaled around 48GB.

\section{Code Availability}
The official implementation of TPDM and pre-trained MRI model checkpoint can be accessed at \url{https://github.com/hyn2028/tpdm}. This repository provides the necessary resources and instructions to replicate the experiments and utilize the TPDM.

\end{document}